\documentstyle[12pt]{article}
\input epsf.tex

\newcommand{\de}{\partial}
\newcommand{\be}{\begin{equation}}
\newcommand{\ee}{\end{equation}}
\newcommand{\bea}{\begin{eqnarray}}
\newcommand{\eea}{\end{eqnarray}}
\newcommand{\bd}{\begin{displaymath}}
\newcommand{\ed}{\end{displaymath}}

\begin{document}
\thispagestyle{empty}
\vskip 1.5cm
\begin{flushright}
ROM2F-96-15
\end{flushright}
\centerline{\large \bf Testing Theories of Gravity with a Spherical} 
\centerline{\large \bf Gravitational Wave Detector}
\vspace{2.2cm}
%%%%%%%%%%%%%%%%%%%%%%%%  AUTORI  %%%%%%%%%%%%%%%%%%%%%%%%%%%%%%
\centerline{\bf M. Bianchi, E. Coccia, C.N. Colacino, 
V. Fafone, F. Fucito}
\vskip 0.5cm
\centerline{\sl Dipartimento di Fisica, 
Universit\`a di Roma II ``Tor Vergata"}  
\centerline{\sl I.N.F.N. \ - \  Sezione di Roma II, }
\centerline{\sl Via Della Ricerca Scientifica \ \ 00133 \ 
Roma \ \ ITALY}
\vskip 2.2cm
\centerline{\large \bf ABSTRACT}
{We consider the possibility of discriminating different theories 
of gravity using a recently proposed gravitational wave detector
of spherical shape. We argue that the spin content of different
theories can be extracted relating the measurements of the excited
spheroidal vibrational eigenmodes to the Newman-Penrose parameters. 
The sphere
toroidal modes cannot be excited by any metric GW and can be thus 
used as a
veto.}
\newpage
The efforts aimed at the detection of  gravitational waves (GW) 
started more than a quarter of century ago and have been,
up to now, unsuccessful \cite{ama, odia}. Resonant bars 
have proved their reliability, being capable of continous
data gathering for long periods of time \cite{asto, joal}. 
Their energy sensitivity
has improved of more than four orders of magnitude since Weber's
pioneering experiment. But a  further improvement is still 
necessary to
achieve successful detection. While further developments of 
bar detectors are under way, two new generations of earth 
based experiments
have been proposed. While detectors based on large laser 
interferometers are already under construction \cite{abbra}, 
resonant detectors of spherical shape are under 
study \cite{odia}.
In the present letter we report on a study about the physical
information that can be obtained thanks to the
spherical symmetry of the latter detectors. More in detail, 
we want
to show that the measurements of the sphere vibrations 
can provide specific information on the field content of the 
gravitational 
theory predicting the observed features of the waves.

We would like to remind the 
reader of the very special position of Einstein's general 
relativity (GR)
among the possible gravitational theories. 
Theories of gravitation, in fact,
can be divided into two families: metric and non-metric 
theories \cite{will}.
The former can be defined to be all theories obeying the following
three postulates:
\begin{itemize}
\item
spacetime is endowed with a metric; 
\item
the world lines of test particles are geodesic of the 
above mentioned
metric;
\item
in local free-falling frames, the non-gravitational 
laws of physics 
are those 
of special relativity.
\end{itemize}
It is an obvious consequence of these postulates 
that a metric 
theory obeys the principle  of equivalence. More 
succintly a theory 
is said 
to be metric if the action of gravitation on the matter sector 
is due exclusively to the metric tensor.
GR is the most famous example of a metric 
theory. Kaluza-Klein type theories, also belong to this
class along with the Brans-Dicke theory. Different 
representatives of 
this class differ for their equations of motion 
which in turn can be 
deduced 
from a lagrangian principle. Up to now non-Einsteinian 
and non-metric 
theories 
\cite{hehl} have been considered a curiosity by the 
majority of 
physicists 
since there seems to be no compelling experimental 
or theoretical 
reason to 
justify their introduction.
This point should perhaps be reconsidered if we think of 
the unique role of string theories in this context. 

String theory seems to lead to a consistent framework 
in which to 
quantize gravity and the other fundamental interactions. 
Gravity emerges from string theory in different forms 
depending on the 
chosen vacuum configuration. Allowed string vacua 
are solutions of the
equations of motion resulting from conformal invariance of the
sigma model which governs the string propagation \cite{witt}. 
The string equations
of motion can also be obtained {\it a posteriori} 
via a variational 
principle. 
Indeed, the elementary excitations of the string with zero mass 
(light sector)
can be described
by an effective lagrangian which is obtained after 
integrating out
the massive (heavy sector) modes of the theory 
in a procedure \`a 
la Wilson. 
 A great variety of string solutions has appeared in 
literature: plane waves, solitons, instantons, rotating and 
non-rotating
black holes. More recently a cosmological solution has appeared
that seems to lead to a potentially detectable quantity 
of gravitational
radiation in a wide spectrum of frequencies \cite{vene}. 
These solutions
can be both metric and non-metric but are certainly 
non-Einsteinian 
since all of these vacua contain
a massless scalar called dilaton. When perturbations 
of these vacua 
are considered all fields fluctuate and the resulting 
theory is non-metric.
But, as in string theory the expectation value of the 
dilaton plays the role
of a coupling constant, it is generally believed that 
some non-perturbative
effect (the most serious candidate at the moment is 
supersymmetry breaking)
gives a mass to the dilaton. If this is true, the dilaton 
interactions with
the other fields may well be negligible \cite{dampol} 
and the theory
may revert to be metric.

These are our motivations for testing theories of gravity 
using the 
measurable 
toroidal and spheroidal vibrational eigenmodes of a 

sphere. The 
signature of 
a non-metric theory could obviously also be detected in 
experiments on the 
equivalence principle.

Before discussing the interaction with an external GW field,
let us consider the basic equations governing the free vibrations
of a perfectly homogeneous, isotropic sphere of radius $R$, 
made of a material having density $\rho$ and Lam\'e coefficients 
$\lambda$ 
and $\mu$ \cite{elast}. 
 	
Following the notation of \cite{lobo}, let $x_{i}, i=1,2,3$ be 
the equilibrium position of the element of the
elastic sphere and $x'_{i}$ be the deformed position then
$u_{i}=x'_{i}-x_{i}$ is the displacement vector. Such vector 
is assumed
small, so that the linear theory of elasticity is applicable. 
The strain 
tensor is defined as $u_{ij}=(1/2)(u_{i,j}+u_{j,i})$
and is related to the stress tensor by
$\sigma_{ij}=\delta_{ij}\lambda u_{ll}+2\mu u_{ij}$.
The equations of motion of the free vibrating sphere are thus
\be
\rho \frac {\de ^{2}u_{i}}{\de  t^{2}}= \frac {\de }{\de 
 x^{j}}(\delta_{ij}\lambda u_{ll}+2\mu u_{ij}) 
\ee
with the boundary condition:
\be
n_{j}\sigma _{ij}=0
\label{contorno}
\ee
at $r=R$ where $n_i\equiv x_{i}/r$ is the unit normal. 
These conditions simply state that the surface of the sphere is 
free to vibrate.
The displacement $u_{i}$ is a time-dependent vector, whose time 
dependence can be factorised as 
$u_{i}({\vec x},t)=u_{i}({\vec x})exp(i\omega t)$,
where $\omega$ is the frequency. The equations of motion 
then become:
\be
\mu \nabla ^{2}{u}_i+(\lambda +\mu )\nabla_i 
(\nabla_j u_j)=-\omega 
^{2}\rho {u}_i 
\ee
Their solutions can be expressed as a sum of a longitudinal and 
two transverse 
vectors:
\be 
\vec{u}(\vec{x})=C_{0}\vec{\nabla}\phi (\vec{x})+C_{1}
\vec{L}\chi (\vec{x})+C_{2}\vec\nabla\times\vec{L}\chi (\vec{x}) 
\label{usol}
\ee
where $C_{0}, C_{1}, C_{2}$ are constants and $\vec{L}
\equiv\vec{x}\times
\vec{\nabla}$ is the angular momentum operator.
Regularity at $r=0$ restricts the scalar functions 
$\phi$ and $\chi$  
to be
expressed as $\phi(r,\theta,\varphi)\equiv j_{l}(qr)Y_{lm}
(\theta ,\varphi)$ and
$\chi(r,\theta,\varphi)\equiv j_{l}(kr)Y_{lm}(\theta ,\varphi)$. 
$Y_{lm}(\theta ,
\varphi)$ are the spherical harmonics and $j_{l}$ the 
spherical Bessel 
functions \cite{jack}: 
\be
j_l(x)=\bigg({1\over x}{d\over dx}\bigg)^l
\bigg(\frac{\sin x}{x}\bigg)
\label{bessel}
\ee
$q^{2}\equiv\rho\omega ^{2}/(\lambda +2\mu)$ and 
$k^{2}\equiv\rho\omega ^{2}/\mu$ are the longitudinal 
and transverse 
wave vectors respectively.

Imposing the boundary conditions (\ref{contorno}) 
at $r=R$ yields 
two families 
of solutions:
\begin{itemize}
\item
{\em Toroidal} modes: these are obtained by setting 
$C_{0}=C_{2}=0$, and $C_{1}
\neq 0$. The eigenfunctions have the form:
\be
\vec{u}^{T}_{nlm}(r,\theta,\varphi)=T_{nl}(r)\vec{L}Y_{lm}
(\theta,\varphi)
\label{paperone} 
\ee
with $T_{nl}(r)$ proportional to $j_{l}(k_{nl}r)$.
The eigenfrequencies are
determined by the boundary conditions (\ref{contorno}) which read
\be
f_{1}(kR)=0
\ee
where 
\be
f_{1}(z)\equiv\frac{d}{dz}\bigg[\frac{j_{l}(z)}{z}\bigg]. 
\ee

\item

{\em Spheroidal} modes: these are obtained by 
setting $C_{1}=0$, $C_{0}\neq 0$ and $C_{2}\neq 0$. 
The eigenfunctions can be conveniently rewritten as
\be
\vec{u}^{S}_{nlm}(\vec{x})=A_{nl}(r)Y_{lm}(\theta ,\varphi)\vec{n}
-B_{nl}(r)\vec{n} \times\vec{L}Y_{lm}(\theta ,\varphi)
\label{quattrostagioni}
\ee
where $A_{nl}(r)$ and $B_{nl}(r)$ are dimensionless radial 
eigenfunctions 
\cite{lobo}, which can be expressed in terms of the
spherical Bessel functions and their derivatives. 
The eigenfrequencies are determined by the boundary conditions 
(\ref{contorno}) which read
\be
det\pmatrix{f_{2}(qR)-{\lambda\over 2\mu}q^{2}R^{2}f_{0}(qR)&l(l+1)
f_{1}(kR)\cr f_{1}(qR)&{1\over 2}f_{2}(kR)+[\frac{l(l+1)}{2}-1]
f_{0}(kR)}=0
\label{minnie}
\ee
where 
\be
f_{0}(z)\equiv\frac{j_{l}(z)}{z^{2}} \quad
f_{2}(z)\equiv\frac{d^{2}}{dz^{2}}j_{l}(z)
\ee
\end{itemize}
The eigenfrequencies can be determined numerically for both 
toroidal and 
spheroidal vibrations. Each mode of order $l$ is $(2l+1)$-fold 
degenerate. 
In the table below we show the value of 
the $(kR)$ roots for the lowest toroidal and spheroidal 
modes of vibration. 
\vskip 1.0cm
\begin{centering}
\begin{tabular}{|c|c|c|c||c|c|c|c|}
\hline
$l$ & $n$ & $(kR)_{toroidal}$ & $(kR)_{spheroidal}$ &
$l$ & $n$ & $(kR)_{toroidal}$ & $(kR)_{spheroidal}$ \\ \hline\hline
0 & 1 & -       & 5.4322 &	3 & 1 & 3.8647  & 3.9489 \\ \hline
  & 2 & -       & 12.138 &	  & 2 & 8.4449  & 6.6959 \\ \hline
  & 3 & -       & 18.492 &	  & 3 & 11.882  & 9.9720 \\ \hline
  & 4 & -       & 24.785 &	  & 4 & 15.175  & 12.900 \\ \hline
1 & 1 & 5.7635  & 3.5895 &	4 & 1 & 5.0946  & 5.0662 \\ \hline
  & 2 & 9.0950  & 7.2306 &	  & 2 & 9.7125  & 8.2994 \\ \hline
  & 3 & 12.323  & 8.4906 &	  & 3 & 13.211  & 11.324 \\ \hline
  & 4 & 15.515  & 10.728 &	  & 4 & 16.544  & 14.467 \\ \hline
2 & 1 & 2.5011  & 2.6497 &	5 & 1 & 6.2658	& 6.1118 \\ \hline
  & 2 & 7.1360  & 5.0878 &	  & 2 & 10.951	& 9.8529 \\ \hline
  & 3 & 10.515  & 8.6168 &	  & 3 & 14.511	& 12.686 \\ \hline
  & 4 & 13.772  & 10.917 &	  & 4 & 17.886	& 15.879 \\ \hline
\end{tabular}
\end{centering}
\vskip .5cm
The eigenfrequency values can be obtained from :
\be
\omega_{nl} = \sqrt{\mu\over\rho} {(kR)_{nl}\over R}
\ee
The detector is assumed to be non-relativistic 
(with sound velocity $v_s\ll c$ and radius $R\ll \lambda$
the GW wavelength) and 
endowed with a high quality factor ($Q_{nl}=\omega_{nl} \tau_{nl}\gg1$, 
where $\tau_{nl}$ is the decay time of the mode $nl$). 
The displacement $\vec u$ of a point in the detector can be 
decomposed in 
normal modes as:
\be
\vec u(\vec{x},t) = \sum_N A_N(t) \vec u_N (\vec x)
\label{expan}
\ee
where $N$ collectively denotes the set of quantum numbers 
identifying the mode.
The basic equation governing the response of the detector is 
\be
\ddot{A}_{N}(t) + \tau _{N}^{-1} \dot{A}_{N}(t) + 
\omega _{N}^{2}A_{N}(t) = 
f_{N}(t)
\label{forzate}
\ee
We assume that the gravitational interaction obeys the 
principle of equivalence 
which has been experimentally supported to high accuracy.
In terms of the 
so-called electric components of the Riemann tensor 
$E_{ij}\equiv R_{0i0j}$,
the driving force $f_N(t)$ is then given by
\be
f_{N}(t) = - M^{-1} E_{ij}(t) \int u_{N}^{i*}(\vec{x}) x^{j}
\rho d^{3}x
\label{forza}
\ee
where $M$ is the sphere mass and we consider the density 
$\rho$ as a constant.
In any metric theory of gravity $E_{ij}$ is a $3\times3$
symmetric tensor, which depends on time, but not on 
spatial components.

Let us now investigate 
which sphere 
eigenmodes can be excited by a metric GW, {\it i.e.} which 
sets of quantum 
numbers N give a non-zero driving force.

a) {\em Toroidal} modes 

\noindent The displacement vector can be expressed as in eq. 
(\ref{paperone}).
Up to a normalisation constant C, the driving force is
\bea
f^{(T)}_{N}(t)&=&-e^{-i\omega_{N}t}\frac{3C}{4\pi 
R^{3}}\int_{0}^{R}dr r^{3}
j_{l}(k_{nl}^{(T)}r)\int_{0}^{\pi}d\theta
\sin\theta\int_{0}^{2\pi}d\phi
\nonumber\\
& &\bigg{\{}\frac{E_{yy}-E_{xx}}{2} \bigg(\sin\theta 
\sin 2\phi\frac{\de  Y_{lm}^{*}}
{\de \theta}+\cos\theta\cos 2\phi\frac{\de  Y_{lm}^{*}}
{\de \phi}\bigg)\nonumber \\
& & \mbox{}+E_{xy}\bigg(\sin\theta\cos 2\phi\frac{\de  
Y_{lm}^{*}}{\de \theta}-
\cos\theta\sin 2\phi\frac{\de  Y_{lm}^{*}}{\de \phi}\bigg)
\nonumber \\
& & \mbox{}+E_{xz}\bigg[-\sin\phi\cos\theta\frac{\de  
Y_{lm}^{*}}{\de \theta}+
(\sin\theta\cos\phi-\frac{\cos^2\theta}{\sin\theta}\cos\phi)
\frac{\de  Y_{lm}^{*}}{\de \phi}\bigg]\nonumber \\
& & \mbox{}+E_{yz}\bigg[\cos\phi\cos\theta\frac{\de 
Y_{lm}^{*}}{\de\theta}
+(\sin\theta\sin\phi-\frac{\cos^2\theta}{\sin\theta}\sin\phi) 
\frac{\de Y_{lm}^{*}}{\de \phi}\bigg]\nonumber \\
& & \mbox{}+\bigg(E_{zz}-\frac{E_{xx}+E_{yy}}{2}\bigg)
\cos\theta\frac{\de  Y_{lm}^{*}}{\de \phi}\bigg\}
\label{intforz}
\eea
Using the equations
\be
\frac{\de  Y_{lm}^{*}}{\de \theta}= (-)^m 
\bigg\lbrack\frac{2l+1}{4\pi}\frac{(l-m)!}{(l+m)!}
\bigg\rbrack^{1\over 2} 
\frac{\de  P_{l}^{m}(\cos\theta)}{\de \theta}e^{-im\phi}
\label{legendre}
\ee
and
\be
\frac{\de  Y_{lm}^{*}}{\de \phi}= -im(-)^m 
\bigg\lbrack\frac{2l+1}{4\pi}\frac{(l-m)!}{(l+m)!}
\bigg\rbrack^{1\over 2}
P_{l}^{m}(\cos\theta)e^{-im\phi}
\label{lavendetta}
\ee
the integration over $\phi$ can be performed. Eq. (\ref{intforz}) 
then contains
integrals over $\theta$ of the form:
\be
\int_0^\pi \bigg[(\sin^2\theta-\cos^2\theta)P_l^{\pm 1}(\cos\theta)
- \sin\theta\cos\theta\frac{\de P_l^{\pm 1}(\cos\theta)}
{\de \theta}\bigg]d\theta
\label{carolina}
\ee
and 
\be
\int_0^\pi \bigg[2\sin\theta\cos\theta P_l^{\pm 2}(\cos\theta)
+ \sin^2\theta\frac{\de P_l^{\pm 2}(\cos\theta)}{\de \theta}
\bigg]d\theta
\label{annamaria}
\ee
After integration by parts, the derivative terms in eqs. 
(\ref{carolina}) and 
(\ref{annamaria}) exactly cancel the non-derivative ones.
The remaining boundary terms vanish too, thanks to the 
periodicity of the trigonometric functions and to the 
regularity of the 
associated Legendre polynomials. 
The vanishing of the above integrals has a profound physical 
consequence.
It means that in any metric theory of gravity the toroidal modes of 
the sphere 
cannot be excited by GW and can thus be used as a veto in the 
detection. In 
this respect we disagree with the results 
of \cite{zhou} that seems to find a zero result only for the 
even $l$ case.
\par
b) {\em Spheroidal} modes \par 
\noindent The forcing term is given by:
\be
f^{(S)}_{N}(t) = - M^{-1} E_{ij}(t) \int x^{j}
\bigg({x^i\over r} A_N (r) 
Y_{lm}(\theta,\varphi) - B_N (r) \epsilon^{ijk} {x_j\over r} L_k 
Y_{lm}(\theta,\varphi)\bigg) \rho d^{3}x
\label{sferforza}
\ee
One is thus lead to compute integrals of the following types
\be
\int x^{j}x^i Y_{lm}(\theta,\varphi) d^3 x
\ee
and
\be
\int x^{j}x^i L_k Y_{lm}(\theta,\varphi) d^3 x
\ee
Since the product $x^i x^j$ can be expressed in terms of the 
spherical
harmonics with $l=0,2$ and the angular momentum operator does 
not change the 
value of $l$, one immediately concludes that in any metric 
theory of gravity
only the $l=0,2$ spheroidal modes of the sphere can be excited. 
At the lowest level there are a total of
five plus one independent spheroidal modes that can be used for 
GW detection
and study.

These results agree with the conclusion of \cite{lobo} and 
generalize
the results of \cite{wago, drei}.

From the analysis of the spheroidal modes active for 
metric GW, we now want
to infer the field content of the theory.
For this purpose it is convenient to express the Riemann tensor in a 
null (Newman-Penrose) tetrad basis \cite{will}.

To lowest non-trivial order in the perturbation the six indipendent 
"electric" components of the Riemann tensor may be expressed in terms
of the Newmann-Penrose (NP) parameters as 
\be
E_{ij} = \pmatrix{-Re\Psi_4-\Phi_{22}&Im\Psi_4&-2\sqrt2 Re\Psi_3\cr 
Im\Psi_4&Re\Psi_4-\Phi_{22}&2\sqrt2 Im\Psi_3\cr 
-2\sqrt2 Re\Psi_3&2\sqrt2 Im\Psi_3&-6\Psi_2}  
\label{newpen}
\ee 
The NP parameters allow the identification of the spin
content of the metric theory responsible for the generation of the wave
\cite{will}. The classification can be summarized in order of incresing
complexity as follows:
\begin{itemize}
\item
General Relativity (spin 2): $\Psi_4\neq 0$ while 
$\Psi_2=\Psi_3=\Phi_{22}=0$.
\item
Tensor-scalar theories (spin 2 and 0): $\Psi_4\neq 0$, $\Psi_3=0$,
$\Psi_2\neq 0$ and/or $\Phi_{22}\neq 0$ ({\it e.g.} Brans-Dicke theory
with $\Psi_4\neq 0$, $\Psi_2=0$, $\Psi_3=0$ and $\Phi_{22}\neq 0$).
\item
Tensor-vector theories (spin 2 and 1):
$\Psi_4\neq 0$, $\Psi_3\neq 0$, $\Phi_{22}=\Psi_2=0$.
\item
Most General Metric Theory (spin 2, 1 and 0): 
$\Psi_4\neq 0$, $\Psi_2\neq 0$, $\Psi_3\neq 0$ and $\Phi_{22}\neq 0$, 
({\it e.g.} Kaluza-Klein theories with $\Psi_4\neq 0$, $\Psi_3\neq 0$, 
$\Phi_{22}\neq 0$ while $\Psi_2=0$).

\end{itemize}

In eq. (\ref{newpen}), we have assumed that the wave comes from a 
localized 
source with wave vector $\vec k$ parallel to the $z$ axis of 
the detector 
frame. 
In this case the NP parameters (and thus the wave polarisation 
states) can be
uniquely determined by monitoring the six lowest spheroidal modes.
If the direction of the incoming wave is not known two 
more unknowns appear in the problem, {\it i.e.} 
the two angles of rotation of the 
detector frame needed to align $\vec k$ along the $z$ direction. 
In order to dispose of this problem one can envisage the 
possibility of 
combining the pieces of information from an array of 
detectors \cite{tinto}.
We restrict our attention to the simplest case in which
the source direction is known. 

The NP parameters of the incident 
wave can be easily obtained from the following argument.
It is well known that any $3\times3$ symmetric tensor such as 
(\ref{newpen})
can be decomposed in the following way
\be
E_{ij}(t)=\sum_{l,m} c_{l,m}(t) S_{ij}^{(l,m)} 
\label{decompo}
\ee
where $S_{ij}^{(0,0)}\equiv \delta_{ij}/\sqrt{4\pi}$
(with $\delta _{ij}$ the Kronecker symbol) and $S_{ij}^{(2,m)}$ 
($m=-2,..2$) 
are five linearly independent symmetric and traceless matrices. 
The following 
explicit representation is particularly suited for exposing the 
spin content
of the GW \cite{lobo}
\bea
S_{ij}^{(0,0)} &=& (\frac{1}{4\pi})^{1\over 2} 
\pmatrix{1&0&0\cr 0&1&0\cr 0&0&1} \quad
S_{ij}^{(2,0)} = (\frac{5}{16\pi})^{1\over 2} 
\pmatrix{-1&0&0\cr 0&-1&0\cr 0&0&2} \nonumber\\\nonumber\\\
S_{ij}^{(2,2)} &=& (\frac{15}{32\pi})^{1\over 2} 
\pmatrix{1&i&0\cr i&-1&0\cr 0&0&0}\quad
S_{ij}^{(2,-2)} = (\frac{15}{32\pi})^{1\over 2} 
\pmatrix{1&-i&0\cr -i&-1&0\cr 0&0&0}\nonumber\\\nonumber\\\
S_{ij}^{(2,1)} &=& -(\frac{15}{32\pi})^{1\over 2} 
\pmatrix{0&0&1\cr 0&0&i\cr 1&i&0}\quad 
S_{ij}^{(2,-1)} = (\frac{15}{32\pi})^{1\over 2} 
\pmatrix{0&0&1\cr 0&0&-i\cr 1&-i&0}
\label{matrici}
\eea
The matrices $S^k$ are trace-orthogonal and are connected to the
spherical harmonics by 
\bea
S_{ij}^{(0,0)}n_in_j&=&Y_{0,0},\qquad S_{ij}^{(2,0)}n_in_j{}=Y_{2,0}
\nonumber\\ 
S_{ij}^{(2,2)}n_in_j&=&Y_{2,2},\qquad S_{ij}^{(2,-2)}n_in_j=Y_{2,-2}
\nonumber\\ 
S_{ij}^{(2,1)}n_in_j&=&Y_{2,1},\qquad S_{ij}^{(2,-1)}n_in_j=Y_{2,-1}
\label{nuovaeq}
\eea
The vector
$n_i$ in eqs. (\ref{nuovaeq}) has been defined after 
eq. (\ref{contorno}).

Taking the scalar product we find
\bea
c_{0,0}(t)&=&\frac{4\pi}{3}S_{ij}^{0,0}E_{ij}(t)\nonumber\\
c_{2,m}(t)&=&\frac{8\pi}{15}S_{ij}^{2,m}E_{ij}(t)\label{coccia}
\eea
For the NP parameters we find
\bea
\Phi_{22} &=& \sqrt{\frac{5}{16\pi}}c_{2,0}(t)
-\sqrt{\frac{1}{4\pi}}c_{0,0}(t)\quad
\Psi_2=-\frac{1}{12}\sqrt{\frac{5}{\pi}}c_{2,0}(t)
-\frac{1}{12}\sqrt{\frac{1}{\pi}}c_{0,0}(t)\cr
Re\Psi_4 &=& -\sqrt{\frac{15}{32\pi}}[c_{2,2}+c_{2,-2}]\quad
Im\Psi_4=-i\sqrt{\frac{15}{32\pi}}[c_{2,2}+c_{2,-2}]\cr
Re\Psi_3 &=& \frac{1}{16}\sqrt{\frac{15}{\pi}}[c_{2,1}-c_{2,-1}]
\quad
Im\Psi_3=\frac{i}{16}\sqrt{\frac{15}{\pi}}[c_{2,1}+c_{2,-1}]
\label{relazionenpc}
\eea
Eqs. (\ref{relazionenpc}) relate
the quantities $c_{l,m}$ with the GW polarization 
states, described
by the NP parameters. A pictorial representation of the six
polarization states connected with the NP parameters is given 
in Figure 1 where the wave vector is assumed parallel
to the $z$ axis as in (\ref{newpen}). 
Starting from the top left corner of the figure
we draw the polarization states $\Psi_2, \Phi_{22}, Re\Psi_4,
Im\Psi_4, Re\Psi_3, Im\Psi_3$.
\vskip .6cm
%%%%%%%%%%%%%%%%%%%%%%% Figura
\centerline{\vbox{\epsfysize=80mm \epsfbox{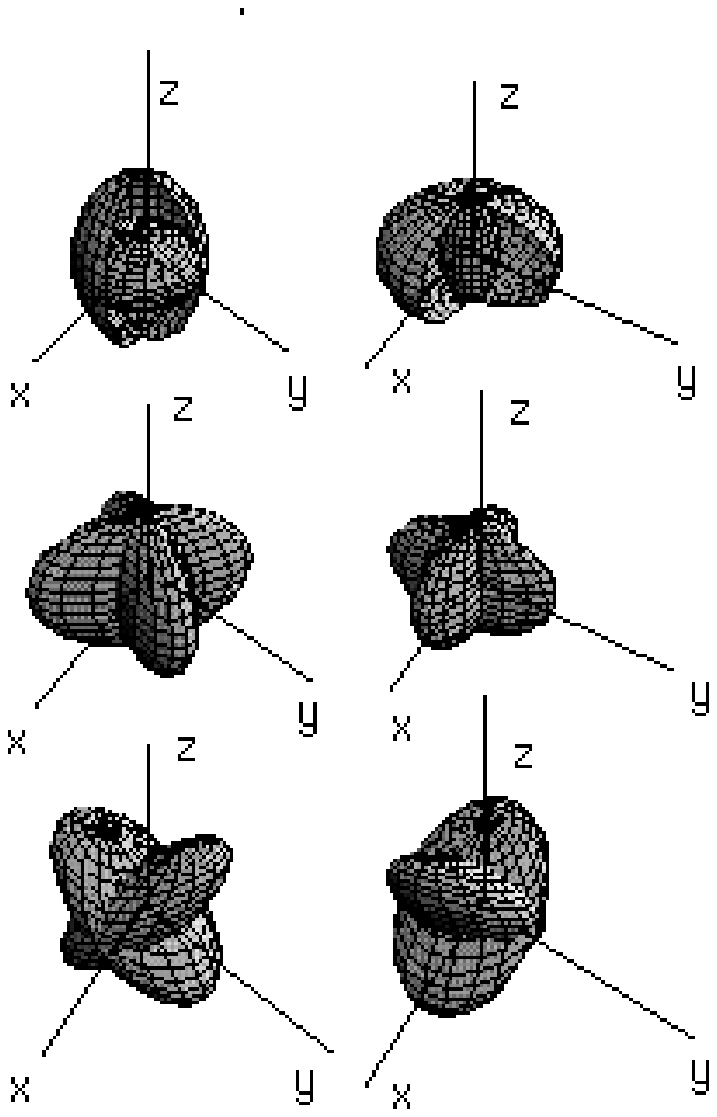}}}
\smallskip
\vskip .4cm
\centerline{\bf Figure 1}
%%%%%%%%%%%%%%%%%%%%%%%%%%%
\vskip .6cm
Eq. (\ref{relazionenpc}) can be put
in correspondence with the output of experimental measurements if 
the $c_{l,m}$ are substituted with their Fourier components at 
the quadrupole and monopole resonant frequencies which, for
the sake of simplicity, we collectively denote by $\omega_0$. 
The $c_{l,m}(\omega_0)$ can be determined in the following way:
once the Fourier amplitudes $A_N(\omega_0)$ are measured, 
by Fourier
transforming (\ref{forzate}) and (\ref{forza}) we get the
Riemann amplitudes $E_{ij}(\omega_0)$ which, using (\ref{coccia}),
yield the desired result.

In order to determine the $A_N(\omega_0)$ amplitudes 
from a given GW signal
two conditions must be fulfilled:
\begin{itemize}
\item
the vibrational states of the five-fold degenerate quadrupole 
and monopole modes must be determined. The quadrupole modes 
can be studied by properly combining
the outputs of a set of at least five motion sensors placed
in independent positions on the 
sphere surface.
Explicit formulas for practical and elegant configurations 
of the motion sensors have been reported by
various authors \cite{zhou, merko, serrano}.
The vibrational state of the monopole mode
is provided directly by the output of any of the above mentioned
motion sensors.
If resonant motion sensors 
are used, since the quadrupole and monopole states 
resonate at different frequencies, a sixth sensor is needed.
\item
The spectrum of the GW signal must be sufficiently 
broadband to overlap with the antenna quadrupole and monopole 
frequencies. 
\end{itemize}
From table 1 we see that the second order quadrupole
spheroidal mode is close to the lowest order monopole and also  
to the toroidal
mode $n=1$, $l=4$. Since it has been demonstrated \cite{clo} that 
the second order 
quadrupole
spheroidal mode has high cross section for general 
relativistic GW (only a
factor 2.6 lower than the cross section of the first order quadrupole 
spheroidal mode), the full analysis of the 
GW signal for the most general metric GW, including
the toroidal mode veto, can be hopefully performed
in a small frequency range. The above reported analysis 
can then be applied
to a large class of GW sources, including
gravitational collapses, inspiralling and coalescing binary systems and 
stochastic sources.

The sensitivity of a multiton spherical detector, 
making use of ultracryogenic 
and superconducting techniques for noise reduction could be such 
to detect 
these events  with a reasonable rate \cite{zhou}.

\vskip 1.5cm
\leftline{\bf\large Acknowledgments}
\vskip .5cm
One of the authors (E.C.) would like to thank J.A.Lobo for 
sharing his insights on the subject.

\end{document}